\documentclass[preprint]{revtex4}

\usepackage{bm}
\usepackage{amsmath}
\usepackage[dvipdfmx]{graphicx}
\usepackage{setspace}
\newcommand{\argmax}{\mathop{\rm arg~max}\limits}

 \newcommand{\sgn}{\mathop{\mathrm{sgn}}}

\begin{document}

\title{Energy landscape analysis of neuroimaging data}

\author{
Takahiro Ezaki$^{1,2}$, Takamitsu Watanabe$^{3}$, Masayuki Ohzeki$^{4}$ and Naoki Masuda$^{5,*}$}

\address{$^{1}$National Institute of Informatics, Hitotsubashi, Chiyoda-ku, Tokyo, Japan\\
$^{2}$JST, ERATO, Kawarabayashi Large Graph Project, c/o Global Research Center for Big Data Mathematics, NII, Chiyoda-ku, Tokyo, Japan\\
$^{3}$Institute of Cognitive Neuroscience, University College London, 17 Queen Square, London, WC1N 3AZ, United Kingdom\\
$^{4}$Graduate School of Information Sciences, Tohoku University, Sendai 980-8579, Japan\\
$^{5}$Department of Engineering Mathematics, University of Bristol, Bristol BS8 1UB, United Kingdom}
\email{naoki.masuda@bristol.ac.uk}



\begin{abstract}
Computational neuroscience models have been used for understanding neural dynamics in the brain and how they may be altered when physiological or other conditions change. We review and develop a data-driven approach to neuroimaging data called the energy landscape analysis. The methods are rooted in statistical physics theory, in particular the Ising model, also known as the (pairwise) maximum entropy model and Boltzmann machine. The methods have been applied to fitting electrophysiological data in neuroscience for a decade, but their use in neuroimaging data is still in its infancy. We first review the methods and discuss some algorithms and technical aspects. Then, we apply the methods to functional magnetic resonance imaging data recorded from healthy individuals to inspect the relationship between the accuracy of fitting, the size of the brain system to be analyzed, and the data length.
\end{abstract}


\maketitle

\section{Introduction}

Altered cognitive function due to various neuropsychiatric disorders seems to result from aberrant neural dynamics in the affected brain \cite{Rolls2011,Uhlhaas2012,Kopell2014,Wang2014a}. Alterations in brain dynamics may also occur in the absence of disorders, in situations such as typical aging, traumatic experiences, emotional responses, and tasks. 
Functional magnetic resonance imaging (fMRI) provides information on the neural dynamics in the brain with a reasonable spatial resolution in a non-invasive manner. There are various analysis methods that can be used to extract the dynamics in neuroimaging data including fMRI signals, such as sliding-window functional connectivity analysis, dynamic causal modeling, oscillation analysis, and biophysical modelling. In the present study, we seek the potential of a different approach: energy landscape analysis.

This method is rooted in statistical physics. The main idea is to map the brain dynamics to the movement of a ``ball'' constrained on an energy landscape inferred from neural data. A ball tends to go downhill and remains near the bottom of a basin in a landscape, whereas it sometimes goes uphill due to random fluctuations that cause it to wander around and possibly transit to another basin (Fig.~\ref{fig:schematic}g). Using the Ising model (equivalent to the Boltzmann machine and the pairwise maximum entropy model (MEM); see \cite{Yeh2010,Schneidman2016} for reviews in neuroscience), we can explicitly construct an energy landscape from multivariate time-series data including fMRI signals recorded at a specified set of regions of interest (ROIs). The pairwise MEM, or, equivalently, the Ising model, has been used to emulate fMRI signals \cite{Fraiman2009, Chialvo2010, Das2014, Hudetz2014,Marinazzo2014}. More recently, we have used the pairwise MEM for fMRI data during rest \cite{Watanabe2013} and sleep \cite{Watanabe2014c} and then developed an energy landscape analysis method and applied it to participants during rest \cite{Watanabe2014b} and during a bistable visual perception task \cite{Watanabe2014a}. In contrast with the aforementioned previous studies \cite{Fraiman2009, Chialvo2010, Das2014, Hudetz2014,Marinazzo2014}, our approach is data driven, with the parameters of the Ising model being inferred from the given data. In the present paper, we review the methods and some technical details. In passing, we introduce new techniques (i.e., different inference algorithms and a Dijkstra-like method). We apply the methods to publicly shared resting-state fMRI data recorded from healthy human participants to validate the new approaches and also to examine the relationship between the accuracy of fit, the size of the brain system (i.e., number of ROIs), and the length of the fMRI data.

\section{Material and methods}

The pipeline of the energy landscape analysis based on the pairwise MEM is illustrated in Fig.~\ref{fig:schematic}.

\subsection{Pairwise maximum entropy model}

\begin{figure}[p]
\centering\includegraphics[width=5.0in]{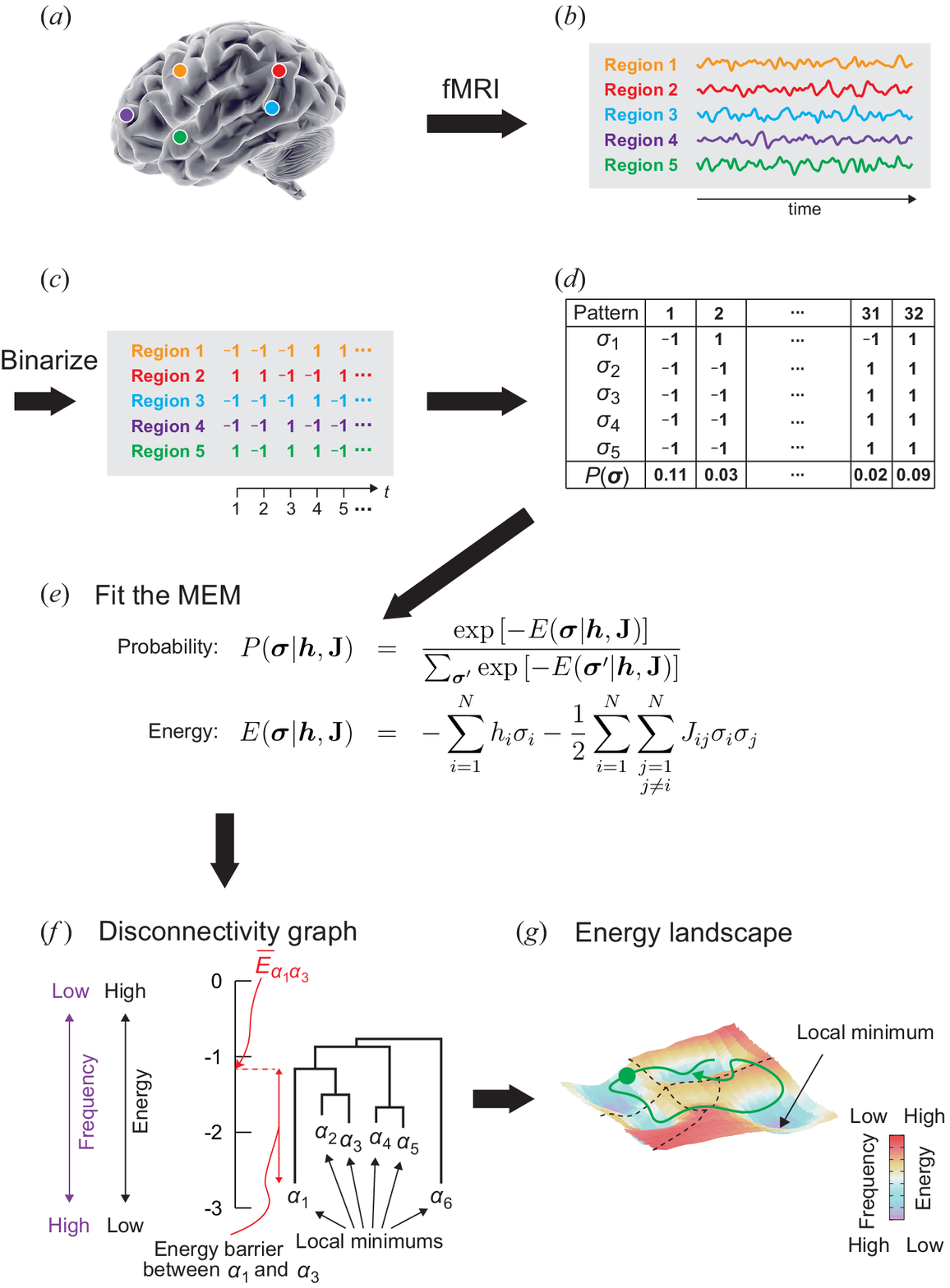}
\caption{Procedures of the energy landscape analysis for fMRI data. (a) ROIs are specified. (b) fMRI signals are measured at the ROIs. (c) The fMRI signal at each ROI and each time point is binarized into $1$ (active) or $-1$ (inactive). 
(d) The normalized frequency is computed for each binarized activity pattern.
(e) The pairwise MEM model (i.e., Boltzmann distribution) is fitted to the empirical distribution of the $2^N$ activity patterns (Eq.~\eqref{eq:Boltzmann}). The energy value is also obtained for each activity pattern (Eq.~\eqref{eq:energy}). (f) Relationships between activity patterns that are energy local minimums are summarized into a disconnectivity graph. 
(g) Schematic of the energy landscape. Each local minimum corresponds to the bottom of a basin. The borders between attractive basins of different local minimums are shown by the dotted curves. Any activity pattern belongs to the basin of a local minimum. Brain dynamics can be interpreted as the motion of a ``ball'' constrained on the energy landscape.
}
\label{fig:schematic}
\end{figure}

First, we specify a brain network of interest (Fig.~\ref{fig:schematic}a) and obtain resting-state (or under other conditions, which are ideally stationary) fMRI signals at the ROIs, resulting in a multivariate time series (Fig.~\ref{fig:schematic}b). We denote the number of ROIs by $N$.

Second, we binarize the fMRI signal at each time point (i.e., in each image volume) and each ROI by thresholding the signal. Then, for each ROI $i$ ($i=1, \ldots, N$), we obtain a sequence of binarized signals representing the brain activity,
$\{\sigma_i (1), \ldots, \sigma_i (t_{\max})\}$, where $t_{\max}$ is the length of the data, $\sigma_i(t)=1$ ($t=1, \ldots, t_{\max}$) indicates that the $i$th ROI is active at time $t$, and $\sigma_i(t)=-1$ indicates that the ROI is inactive (Fig.~\ref{fig:schematic}c). The threshold is arbitrary, and we set it to the time average of $\sigma_i(t)$ for each $i$. The activity pattern of the entire network at time $t$ is given by an $N$-dimensional vector $\bm{\sigma} \equiv (\sigma_1, \ldots,\sigma_N) \in \{-1,1\}^N$, where we have suppressed $t$. Note that there are $2^N$ possible activity patterns in total. Binarization is not readily justified given continuously distributed fMRI signals. However, we previously showed that the pairwise MEM with binarized signals predicted anatomical connectivity of the brain better than other functional connectivity methods that are based on non-binarized continuous fMRI signals and that ternary as opposed to binary quantization did not help to improve the results \cite{Watanabe2013}.

Third, we calculate the relative frequency with which each activity pattern is visited, $P_{\rm{empirical}}(\bm{\sigma})$
(Fig.~\ref{fig:schematic}d). To $P_{\rm{empirical}}(\bm{\sigma})$, we fit the Boltzmann distribution given by
\begin{equation}
P(\bm{\sigma}|\bm{h},\mathbf{J}) = \frac{\exp{\left[-E(\bm{\sigma}|\bm{h},\mathbf{J})\right]}}{\sum_{\bm{\sigma}^{\prime}}\exp{\left[-E(\bm{\sigma}^{\prime} | \bm{h},\mathbf{J})\right]}}
\label{eq:Boltzmann},
\end{equation}
where
\begin{equation}
E(\bm{\sigma}|\bm{h},\mathbf{J}) = - \sum_{i=1}^N h_i \sigma_i - \frac{1}{2}\sum_{i=1}^N\sum_{\substack{j=1\\ j\neq i}}^NJ_{ij}\sigma_i \sigma_j,
\label{eq:energy}
\end{equation}
is the energy, and $\bm{h}= \{h_i\}$ and $\mathbf{J} = \{J_{ij}\}$ ($i,j =1, \ldots, N$) are the parameters of the model (Fig.~\ref{fig:schematic}e). We assume $J_{ij} = J_{ji}$ and $J_{ii}=0$ $(i, j = 1, \ldots, N)$. The principle of maximum entropy imposes that we select $\bm h$ and $\mathbf{J}$ such that $\langle \sigma_i \rangle_{\rm{empirical}} = \langle \sigma_i \rangle_{\rm{model}}$ and $\langle \sigma_i\sigma_j \rangle_{\rm{empirical}} = \langle \sigma_i\sigma_j \rangle_{\rm{model}}$ $(i, j = 1, \ldots, N)$, where $\langle \cdots \rangle_{\rm{empirical}}$ and $\langle \cdots \rangle_{\rm{model}}$ represent the mean with respect to the empirical distribution (Fig.~\ref{fig:schematic}d) and the model distribution (Eq.~\eqref{eq:Boltzmann}), respectively. By maximizing the entropy of the distribution under these constraints, we obtain the Boltzmann distribution given by Eq.~\eqref{eq:Boltzmann}.
Some algorithms for the fitting will be explained in section~\ref{sec:algorithm}. Equation~\eqref{eq:Boltzmann} indicates that an activity pattern with a high energy value is not frequently visited and vice versa. 
Values of $h_i$ and $J_{ij}$ represent the baseline activity at the $i$th ROI and the interaction between the $i$th and $j$th ROIs, respectively. 
Equation~\eqref{eq:energy} implies that, if $h_i$ is large, the energy is smaller with $\sigma_i=1$ than with $\sigma_i=-1$, such that the $i$th ROI tends to be active.

\subsection{Algorithms to estimate the pairwise MEM}\label{sec:algorithm}

In this section, we review three algorithms to estimate the parameters of the MEM, i.e., $\bm h$ and $\mathbf{J}$.

\subsubsection{Likelihood maximization\label{sub:max likelihood}}

In the maximum likelihood method, we solve
\begin{equation}
(\bm{h},\mathbf{J}) = \argmax_{\bm{h},\mathbf{J}} { \mathcal{L}(\bm{h},\mathbf{J})},
\end{equation}
where  $\mathcal{L}(\bm{h},\mathbf{J})$ is the likelihood given by
\begin{equation}
\mathcal{L}(\bm{h},\mathbf{J}) = \prod_{t=1}^{t_{\rm{max}}} P(\bm{\sigma}(t) | \bm{h},\mathbf{J}).
\end{equation}
We can maximize the likelihood by a gradient ascent scheme 
\begin{eqnarray}
h_{i}^{\rm{new}} - h_{i}^{\rm{old}}&=&  \frac{\epsilon}{t_{\rm{max}}} \frac{\partial}{\partial h_{i}} \log \mathcal{L}(\bm{h},\mathbf{J}) =   \epsilon \left(\langle \sigma_i \rangle_{\rm{empirical}} - \langle \sigma_i \rangle_{\rm{model}} \right) ,
\label{eq:h_i update gradient ascent}\\
J_{ij}^{\rm{new}} - J_{ij}^{\rm{old}} &=&  \frac{\epsilon}{t_{\rm{max}}} \frac{\partial}{\partial J_{ij}} \log \mathcal{L}(\bm{h},\mathbf{J}) =  \epsilon \left(\langle \sigma_i\sigma_j \rangle_{\rm{empirical}} - \langle \sigma_i\sigma_j \rangle_{\rm{model}} \right),
\label{eq:J_ij update gradient ascent}
\end{eqnarray}
where the superscripts new and old represent the values after and before a single updating step, respectively, and $\epsilon$ $(>0)$ is a constant. A slightly different updating scheme called the iterative scaling algorithm \cite{Darroch1972}, where the right-hand side of Eqs.~\eqref{eq:h_i update gradient ascent} and \eqref{eq:J_ij update gradient ascent} is replaced by $\epsilon \sgn(\langle \sigma_i\rangle)\log (\langle \sigma_i\rangle_{\rm{empirical}}/\langle \sigma_i\rangle_{\rm{model}})$ and $\epsilon \sgn(\langle \sigma_i\sigma_j\rangle)\log (\langle \sigma_i\sigma_j\rangle_{\rm{empirical}}/\langle \sigma_i\sigma_j\rangle_{\rm{model}})$, respectively, is sometimes used as well \cite{Schneidman2006,Tang2008,Yeh2010,Watanabe2013}.
Because Eq.~\eqref{eq:Boltzmann} is concave in terms of $\bm{h}$ and $\mathbf{J}$ (which we can show by verifying that the Hessian of $\log \mathcal{L}$ is a type of sign-flipped covariance matrix, which is negative semi-definite), the gradient ascent scheme yields the maximum likelihood estimator.
%
%
Because a single updating step involves all the $2^N$ activity patterns to calculate $\langle \sigma_i \rangle_{\rm{model}}$ and $\langle \sigma_i\sigma_j \rangle_{\rm{model}}$,
likelihood maximization is computationally costly for large $N$.

Our Matlab code to calculate the maximum likelihood estimator for arbitrary multivariate time-series data is available in the electronic supplementary material.

\subsubsection{Pseudo-likelihood maximization}

The pseudo-likelihood maximization method approximates the likelihood function as follows:
\begin{equation}
\mathcal{L}(\bm{h},\mathbf{J}) \approx \prod_{t=1}^{t_{\rm{max}}} \prod_{i=1}^{N} \tilde{P}(\sigma_i|\bm{h},\mathbf{J},\bm{\sigma}_{/i}(t)),\label{eq:pseudo-likelihood}
\end{equation}
where $\tilde{P}$ represents the Boltzmann distribution for a single spin, $\sigma_i$, given the other $\sigma_j (j\neq i)$ values fixed to $\bm{\sigma}_{/i}(t) \equiv (\sigma_1(t), \ldots, \sigma_{i-1}(t), \sigma_{i+1}(t), \ldots, \sigma_N(t))$ \cite{Besag1975}. In other words,
\begin{equation}
\tilde{P}(\sigma_i|\bm{h},\mathbf{J},\bm{\sigma}_{/i}(t)) = \frac{\displaystyle\exp\left[h_i\sigma_i + \sum_{{\substack{j=1\\ j\neq i}}}^N J_{ij} \sigma_i \sigma_j (t)\right]}
{\displaystyle\sum_{\sigma_i^{\prime} = -1,+1} \exp\left[h_i\sigma_i^{\prime} + \sum_{{\substack{j=1\\ j\neq i}}}^N J_{ij} \sigma_i^{\prime} \sigma_j(t)\right]}.
\label{eq:pseudo}
\end{equation}
We call the right-hand side of Eq. \eqref{eq:pseudo-likelihood} the pseudo-likelihood. Although this is a mean-field approximation neglecting the influence of $\sigma_i$ on $\sigma_j$ ($j\neq i$), the estimator obtained by the maximization of the pseudo-likelihood approaches the maximum likelihood estimator as $t_{\max}\rightarrow \infty$ \cite{Besag1975}. A gradient ascent updating scheme is given by
\begin{eqnarray}
h_{i}^{\rm{new}} - h_{i}^{\rm{old}}&=&   \epsilon \left(\langle \sigma_i \rangle_{\rm{empirical}} - \langle \sigma_i \rangle_{\tilde{P}}\right),\\
J_{ij}^{\rm{new}} - J_{ij}^{\rm{old}} &=&   \epsilon \left(\langle \sigma_i\sigma_j \rangle_{\rm{empirical}} - \langle \sigma_i\sigma_j \rangle_{\tilde{P}} \right),
\end{eqnarray}
where $\langle{\sigma_i}\rangle_{\tilde{P}}$ and $\langle{\sigma_i\sigma_j}\rangle_{\tilde{P}}$ are the mean and correlation with respect to distribution $\tilde{P}$ (Eq.~\eqref{eq:pseudo}) and are given by
\begin{equation}
\langle{\sigma_i}\rangle_{\tilde{P}} = \frac{1}{t_{\rm{max}}}\sum_{t=1}^{t_{\rm{max}}} \tanh \left[h_i + \sum_{\substack{j'=1\\ j'\neq i}}^N J_{ij^{\prime}}\sigma_{j^{\prime}}(t)\right]
\end{equation}
and
\begin{equation}
\langle{\sigma_i\sigma_j}\rangle_{\tilde{P}} = \frac{1}{t_{\rm{max}}}\sum_{t=1}^{t_{\rm{max}}} \sigma_j(t)\tanh \left[h_i + \sum_{\substack{j'=1\\ j'\neq i}}^N J_{ij^{\prime}}\sigma_{j^{\prime}}(t)\right],
\end{equation}
respectively. It should be noted that this updating rule circumvents the calculation of  $\langle \sigma_i \rangle_{\rm{model}}$ and $\langle \sigma_i\sigma_j \rangle_{\rm{model}}$, which the gradient ascent method to maximize the original likelihood uses and involves $2^N$ terms.

\subsubsection{Minimum probability flow}

Different from the likelihood and pseudo-likelihood maximization, the minimum probability flow method \cite{Sohl-Dickstein2011} is not based on the likelihood function. Consider relaxation dynamics of a probability distribution, $P(\bm{\sigma};\tau)$, on the $2^N$ activity patterns whose master equation is given by
\begin{equation}
\frac{dP(\bm{\sigma};\tau)}{d\tau} = \sum_{\bm{\sigma}'}\left[W(\bm{\sigma}|\bm{\sigma}')P(\bm{\sigma}';\tau) - W(\bm{\sigma}'|\bm{\sigma})P(\bm{\sigma};\tau)\right],
\end{equation}
where $W(\bm{\sigma}|\bm{\sigma}')$ is a transition rate from activity pattern $\bm{\sigma}'$ to activity pattern $\bm{\sigma}$. 
As the initial condition, we impose $P(\bm{\sigma};0) = P_{\rm{empirical}}(\bm{\sigma})$. By choosing 
\begin{equation}
W(\bm{\sigma}|\bm{\sigma}') = 
\begin{cases}
		\exp\left[- \frac{1}{2}(E(\bm{\sigma}|\bm{h},\mathbf{J})- E(\bm{\sigma}'|\bm{h},\mathbf{J}))\right] & (\bm{\sigma} \text{ and } \bm{\sigma}^{\prime} \text{ are neighboring patterns}),\\
	0  & (\text{otherwise}), 
\end{cases}
\end{equation}
where $\bm{\sigma}$ and $\bm{\sigma}^{\prime}$ are neighbors if they are only different at one ROI, we obtain a standard Markov chain Monte Carlo method such that
$P(\bm{\sigma};\tau)$ converges to the Boltzmann distribution given by Eq.~\eqref{eq:Boltzmann}.

In the minimum probability flow method, we look for $\bm{h}$ and $\mathbf{J}$ values for which $P(\bm{\sigma};\tau)$ changes little in the relaxation dynamics at $\tau=0$ \cite{Sohl-Dickstein2011}.
The idea is that only a small amount of relaxation is necessary if the initial distribution, i.e., 
$P(\bm{\sigma};0) (= P_{\rm{empirical}}(\bm{\sigma}))$, is sufficiently close to the equilibrium distribution, i.e., the Boltzmann distribution. 
The Kullback-Leibler (KL) divergence between the empirical distribution, $P(\bm{\sigma};0)$, and a probability distribution after an infinitely small relaxation time, $P(\bm{\sigma};\epsilon)$, is 
approximated as 
\begin{eqnarray}
D_{\rm{KL}}(P(\bm{\sigma};0) \| P(\bm{\sigma};\epsilon)) & \approx&  \left. D_{\rm{KL}}(P(\bm{\sigma};0) \| P(\bm{\sigma};\tau))\right|_{\tau=0} + \epsilon \left.\frac{d D_{\rm{KL}}(P(\bm\sigma;0) \| P(\bm\sigma;\tau))}{d\tau}\right|_{\tau=0}\nonumber\\
&=& \left.\epsilon \frac{d}{d\tau} \left[\sum_{\bm{\sigma}\in \mathcal{D}}P(\bm{\sigma};0)\log\frac{P(\bm{\sigma};0)}{P(\bm{\sigma};\tau)} \right]\right|_{\tau=0}=-\left. \epsilon \sum_{\bm{\sigma}\in \mathcal{D}}\frac{dP(\bm{\sigma};\tau)}{d\tau} \right|_{\tau=0} \nonumber\\
&= &\frac{\epsilon}{t_{\rm{max}}} \sum_{t=1}^{t_{\rm{max}}}  \sum_{\bm{\sigma}' \in \Omega \backslash\mathcal{D}}  W(\bm{\sigma}'|\bm{\sigma}(t)),
\label{eq:cost minimum probability flow}
\end{eqnarray}
where $D_{\rm{KL}}(P(\bm{\sigma};0) \| P(\bm{\sigma};\epsilon)) =\sum_{\bm{\sigma}\in\Omega}P(\bm{\sigma}; 0)\log\left[P(\bm{\sigma}; 0)/P(\bm{\sigma}; \epsilon)\right]$ is the KL divergence, which quantifies the discrepancy between two distributions, $\Omega$ is the set of all the $2^N$ activity patterns, and $\mathcal{D} = \{\bm{\sigma}\in \Omega|P_{\rm{empirical}} (\bm{\sigma})>0 \}$ is a set of activity patterns that appear at least once in the empirical data.
The minimum probability flow method minimizes the last quantity in Eq.~\eqref{eq:cost minimum probability flow}, i.e., the probability flow from activity patterns that appear in the data but not those that do not.
Therefore, the method is not effective when $N$ is small and $t_{\max}$ is large such that most activity patterns appear in the data.
However, when $N$ is large or $t_{\max}$ is small, this algorithm is efficient in terms of the computation time and memory space \cite{Sohl-Dickstein2011}.
A gradient descent method on 
$D_{\rm{KL}}(P(\bm{\sigma};0) \| P(\bm{\sigma};\epsilon))$ is practically used for determining $\bm h$ and $\mathbf{J}$.

\subsection{Accuracy indices}

Fully describing an empirical distribution requires $2^N-1$ parameters, whereas the pairwise MEM only uses $N +N(N-1)/2$ parameters.
The pairwise MEM imposes that the first two moments of $\sigma_i$ agree between the empirical data and the model. However, the model may be inaccurate in describing higher-order correlations in the empirical data. Most previous studies used one or both of the following two measures to quantify the accuracy with which the MEM fitted the empirical data.

The first measure is defined by
\begin{equation}
\frac{I_2}{I_N}=\frac{S_1-S_2}{S_1-S_{N}},
\label{eq:accuracy_e}
\end{equation}
which ranges between 0 and 1 for the maximum likelihood estimator, and $S_k\equiv - \sum_{\bm{\sigma}\in\Omega}P_k(\bm{\sigma})\log P_k(\bm{\sigma})$ is the Shannon entropy of the maximum entropy model incorporating correlations up to the $k$th order \cite{Schneidman2003,Schneidman2006}. The so-called independent MEM, in which we suppress any interaction between the elements (i.e., $J_{ij}=0$ for $i,j=1,\ldots,N$), gives $P_1(\bm\sigma)$. The pairwise MEM gives
$P_2(\bm\sigma)$. The empirical distribution (i.e., $P_{\rm{empirical}}(\bm\sigma)$) is identical to $P_N(\bm\sigma)$.
The denominator of Eq. \eqref{eq:accuracy_e}, $S_1-S_N \equiv I_N$, is referred to as the multi-information, which quantifies the total contribution of the second or higher order correlation to the entropy of the empirical distribution. The numerator, $S_1-S_2 \equiv I_2$, is equal to the contribution of the pairwise correlation. If $I_2/I_N = 1$, the pairwise correlation alone accounts for all the correlations present in the empirical data. If $I_2/I_N = 0$, the pairwise correlation does not deliver any information.

The second measure is defined by
\begin{equation}
r = \frac{D_{\rm{KL}} (P_1(\bm{\sigma}) \| P_N(\bm{\sigma}))-D_{\rm{KL}} (P_2(\bm{\sigma}) \|P_N(\bm{\sigma}))}{D_{\rm{KL}} (P_1(\bm{\sigma}) \| P_N(\bm{\sigma}))}.\label{eq:accuracy}
\end{equation}
Note that $r$ also ranges between 0 and 1 for the maximum likelihood estimator \cite{Shlens2006, Yeh2010, Watanabe2013}. If the pairwise MEM produces a distribution closer to the empirical distribution than the independent MEM does, $r$ is large. If the pairwise MEM and the independent MEM are similar in terms of the proximity to the empirical distribution, we obtain $r\approx 0$.
For the maximum likelihood estimator, we obtain $I_2/I_N = r$ \cite{Tang2008,Yeh2010}.

\subsection{Disconnectivity graph and energy landscape}

Once we have estimated the pairwise MEM, we construct a dendrogram referred to as a disconnectivity graph \cite{Becker1997}, as shown in Fig.~\ref{fig:schematic}f. In the disconnectivity graph, a leaf (with a loose end open downwards) corresponds to an activity pattern $\bm\sigma$ that is a local minimum of the energy, i.e., an activity pattern whose frequency is higher than any other activity pattern in the neighborhood of $\bm\sigma$. The neighborhood of $\bm\sigma$ is defined as the set of the $N$ activity patterns that are different from $\bm\sigma$ only at one ROI. In the disconnectivity graph, the vertical position of the endpoint of the leaf represents its energy value, which specifies the frequency of appearance, with a lower position corresponding to a higher frequency. The branching structure of the disconnectivity graph describes the energy barrier between any pair of activity patterns that are local minimums. For example, to transit from local minimum ${\bm\alpha_1}$ to local minimum ${\bm \alpha_3}$ in Fig.~\ref{fig:schematic}f, the brain dynamics have to overcome the height of the energy barrier (shown by the double-headed arrow). If the barrier is high, transitions between the two activity patterns occur with a low frequency.

The disconnectivity graph is obtained by the following procedures.
First, we enumerate local minimums, i.e., the activity patterns whose energy is smaller than that of all neighbors. Then, for a given pair of local minimums ${\bm\alpha}$ and ${\bm\alpha'}$, we consider a path connecting them, ${\bm \alpha}\leftrightarrow {\bm \alpha'}$, where a path is a sequence of activity patterns starting from ${\bm \alpha}$ and ending at ${\bm \alpha'}$ such that any two consecutive activity patterns on the path are neighboring patterns. We denote by $E_{\max}(\bm\alpha \leftrightarrow \bm\alpha')$ the largest energy value among the activity patterns on path $\bm\alpha \leftrightarrow \bm\alpha'$. The brain dynamics on this path must climb up the hill to go through the activity pattern with energy $E_{\max}(\bm\alpha \leftrightarrow \bm\alpha')$ to travel between $\bm\alpha$ and $\bm\alpha'$. Because a large energy value corresponds to a low frequency of the activity pattern, a large $E_{\rm{max}}(\bm\alpha \leftrightarrow \bm\alpha')$ value implies that the frequency of switching between $\bm\alpha$ and $\bm\alpha'$ along this path is low. Because various paths may connect $\bm\alpha$ and $\bm\alpha'$, we consider
\begin{equation}
\overline{E}_{\bm\alpha\bm\alpha'} = \min_{\bm\alpha\leftrightarrow \bm\alpha'} E_{\max}(\bm\alpha\leftrightarrow\bm\alpha').
\end{equation}
If we remove all the rarest activity patterns whose energy is equal to or larger than $\overline{E}_{\bm\alpha\bm\alpha'}$, $\bm\alpha$ and $\bm\alpha'$ are disconnected (i.e., no path connecting them exists).
The energy barrier for the transition from $\bm\alpha$ to $\bm\alpha'$ is given by $\overline{E}_{\bm\alpha\bm\alpha'}-E(\bm\alpha)$.

To calculate $\overline{E}_{\bm\alpha\bm\alpha'}$, we employ a Dijkstra-like method as follows. Consider the hypercube composed of $2^N$ activity patterns. By definition, two nodes (i.e., activity patterns) are adjacent to each other (i.e., directly connected by a link) if they are neighboring activity patterns. Each node has degree (i.e., number of neighbors) $N$. Then, fix a local minimum activity pattern $\bm\alpha$
and look for $\overline{E}_{\bm\alpha\bm\alpha'}$ for all local minimums $\bm\alpha'$.
We set $\overline{E}_{\bm\alpha\bm\alpha} = E(\bm\alpha)$ and $\overline{E}_{\bm\alpha\bm\alpha'} = E(\bm\alpha')$ for all $\bm\alpha'$ that are neighbors of $\bm\alpha$. These values are finalized and will not be changed. 
$\overline{E}_{\bm\alpha\bm\alpha'}$ for the other $2^N-N-1$ local minimums $\bm\alpha'$ are initialized to $\infty$. Then, we iterate the following procedures until $\overline{E}_{\bm\alpha\bm\alpha'}$ values for all the nodes $\bm\alpha'$ are finalized. (i) For each finalized $\bm\alpha'$, update $E_{\bm\alpha\bm\alpha'}$ for its all unfinalized neighbors $\bm\alpha''$ using
\begin{equation}
\overline{E}_{\bm\alpha\bm\alpha''}^{\rm{new}} = \left\{ \begin{array}{ll}
\min \{\overline{E}_{\bm\alpha\bm\alpha''}^{\rm{old}}, \overline{E}_{\bm\alpha\bm\alpha'}\} & (\overline{E}_{\bm\alpha\bm\alpha'} \geq E(\bm{\alpha''}), \\
E(\bm{\alpha''}) & (\overline{E}_{\bm\alpha\bm\alpha'} < E(\bm{\alpha''}).\\
\end{array} \right.
\end{equation}
(ii) Find $\bm\alpha'$ with the smallest unfinalized $\overline{E}_{\bm\alpha\bm\alpha'}$ value and finalize it.
(iii) Repeat steps (i) and (ii). If we carry out the entire procedure for each local minimum $\bm\alpha$, we obtain $\overline{E}_{\bm\alpha\bm\alpha'}$ for all pairs of local minimums.

By collecting pairs of local minimums that have the same $\overline{E}_{\bm\alpha\bm\alpha'}$ value, we specify a set of local minimums that should be located under the same branch. This information is sufficient for drawing the dendrogram of local minimums, i.e., the disconnectivity graph.

Each local minimum has a basin of attraction in the state space, $\Omega$. Each activity pattern, denoted by $\bm\sigma$, usually belongs to one of the attractive basins, which is determined as follows.
(i) Unless $\bm\sigma$ is a local minimum, move to the neighboring activity pattern that has the smallest energy value. (ii) Repeat step (i) until a local minimum, denoted by $\bm\alpha$, is reached. We conclude that $\bm\sigma$ belongs to the attractive basin of $\bm\alpha$. (iii) Repeat steps (i) and (ii) for all the initial activity patterns $\bm\sigma\in \Omega$. 

Using the information on the local minimums and attractive basins, the dynamics of the activity pattern are illustrated as the motion of a ``ball'' on the energy landscape, as schematically shown in Fig.~\ref{fig:schematic}g as a hypothetical two-dimensional landscape. The local minimums and energy barriers in Fig.~\ref{fig:schematic}g correspond to those shown in the disconnectivity graph (Fig.~\ref{fig:schematic}f). 

\subsection{0/1 versus 1/-1}

We remark on two binarization schemes. In statistical physics, the pairwise MEM, or the Ising model, usually employs
$\sigma_i \in \{-1,1\}~(i=1,\ldots,N)$ rather than $\tilde{\sigma}_i \in \{0,1\}$. The former convention respects the symmetry between the two spin states and is also convenient in some analytical calculations of the model that exploit the relationship $(\sigma_i)^2 =1$ regardless of $\sigma_i$ \cite{Mezard1987, Nishimori2001}. For neuronal spike data, $\tilde{\sigma}_i \in \{0,1\}$ is often used \cite{Shlens2006,Roudi2009,Ganmor2011a}, whereas $\sigma_i \in \{-1,1\}$ is also commonly used \cite{Schneidman2006,Tang2008,Yu2008,Roudi2009a}. 
For fMRI data, our previous work employed $\tilde{\sigma}_i\in\{0,1\}$ \cite{Watanabe2013,Watanabe2014a,Watanabe2014b,Watanabe2014c}.
The use of $\tilde{\sigma}_i\in\{0,1\}$ in representing neuronal spike trains has a rationale in being able to express the instantaneous firing rate in a simple form $\sum_{i=1}^N \sigma_i$ and synchronous firing of neurons by a simple multiplication \cite{Amari2003}. For example, three neurons simultaneously fire if and only if $\sigma_1\sigma_2\sigma_3=1$. It should also be noted that the iterative scaling algorithm for maximizing the likelihood (section~\ref{sub:max likelihood})) does not generally work for $\sigma_i \in \{-1,1\}$ because $\langle \sigma_i\rangle_{\rm{empirical}}$ and $\langle \sigma_i\rangle_{\rm{model}}$, the logarithm of whose ratio is used in the algorithm, may have opposite signs.

The energy in the case of $\tilde{\sigma}_i \in \{0,1\}$ is defined as 
\begin{equation}
\tilde{E}(\tilde{\bm{\sigma}}|\tilde{\bm{h}},\tilde{\mathbf{J}}) = - \sum_{i=1}^N \tilde{h}_i\tilde{\sigma}_i  - \frac{1}{2}\sum_{i=1}^N\sum_{\substack{j=1\\j\neq i}}^N\tilde{J}_{ij}\tilde{\sigma}_i\tilde{\sigma}_j.
\end{equation}
Mathematically, the two representations are equivalent to the one-to-one relationship, $2 \tilde{\sigma}_i -1 = \sigma_i~(i=1,\ldots,N)$, which results in
\begin{eqnarray}
\tilde{h}_i &=& 2h_i - 2 \sum_{{\substack{j=1\\ j\neq i}}}^NJ_{ij},\\
\tilde{J}_{ij} &=& 4J_{ij}. 
\end{eqnarray}


\section{Results}
\subsection{Accuracy of the three methods}
We applied the three methods to estimate the pairwise MEM to
resting-state fMRI signals recorded from two healthy adult individuals in the Human Connectome Project. We extracted ROIs from three brain systems, i.e., default mode network (DMN, $N_{\rm{ROI}}=12$), fronto-parietal network (FPN, $N_{\rm{ROI}}=11$), and cingulo-opercular network (CON, $N_{\rm{ROI}}=7$), using the ROIs whose coordinates were identified previously \cite{Fair2009}. We had $t_{\max} = 9560$ volumes in total.

\begin{figure}[p]
\centering\includegraphics[width=5.0in]{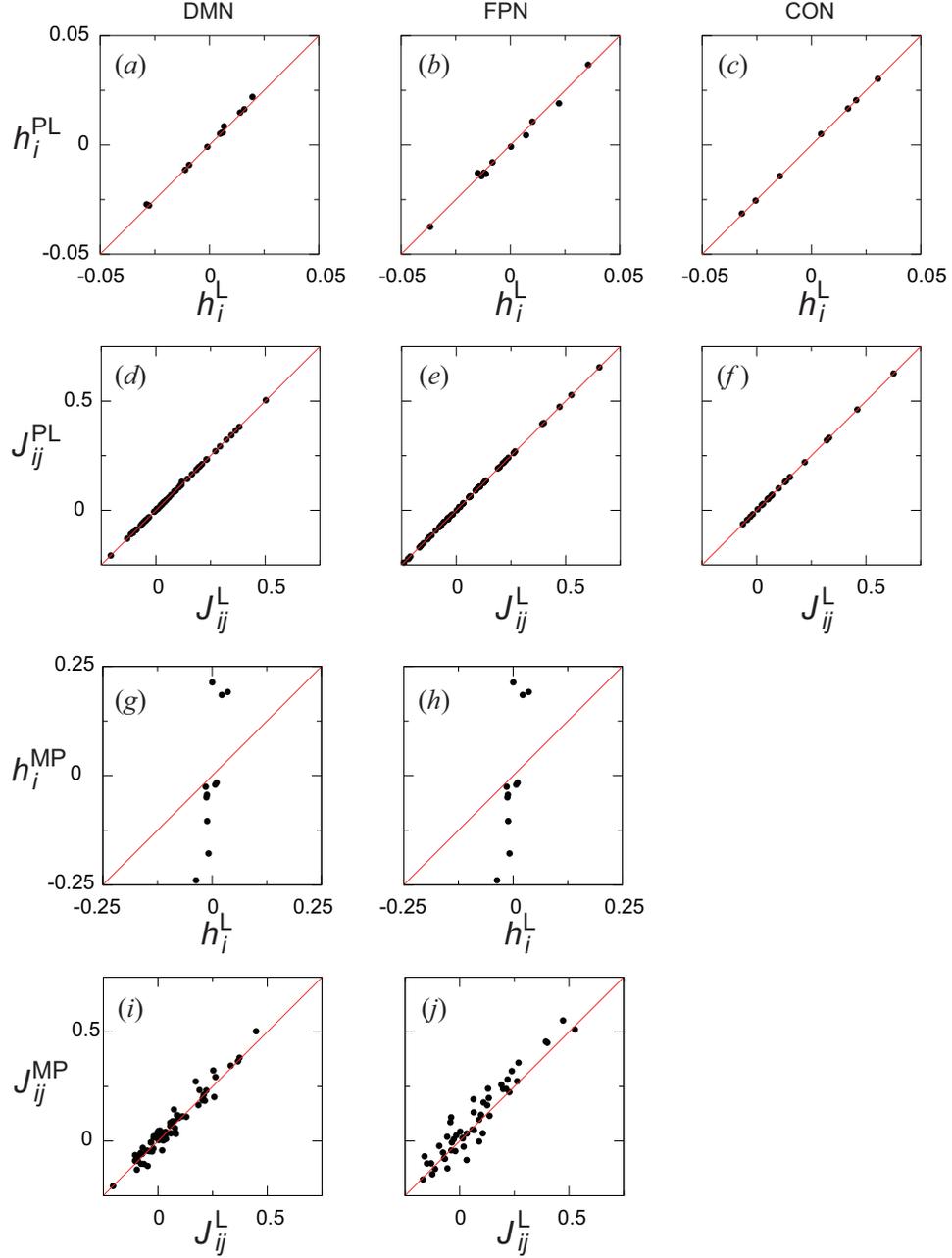}
\caption{Estimated parameter values compared between the different methods. (a)--(f): Comparison between the likelihood maximization and the pseudo-likelihood estimation. (g)--(j): Comparison between the likelihood maximization and the minimum probability flow method. (a), (d), (g), (i): DMN. (b), (e), (h), (j): FPN. (c), (f): CON. Superscripts L, PL, and MP stand for likelihood maximization, pseudo-likelihood maximization, and the minimum probability flow, respectively.}
\label{fig:comparison}
\end{figure}

The estimated parameter values are compared between likelihood maximization and pseudo-likelihood maximization in Fig.~\ref{fig:comparison}a--f. For all the networks, the results obtained by the pseudo-likelihood maximization are close to those obtained by the likelihood maximization, in particular for $\mathbf{J}$. The results obtained by the likelihood maximization and those obtained by the minimum probability flow are compared in Fig.~\ref{fig:comparison}g--j for the DMN and FPN. We did not apply the minimum probability flow method to the CON because all of the $2^8$ activity patterns appeared at least once, i.e., $\mathcal{D} = \Omega$, which made the right-hand side of Eq.~\eqref{eq:cost minimum probability flow} zero. The figure indicates that the estimation by the minimum probability flow deviates from that by the likelihood maximization more than the estimation by the pseudo-likelihood maximization does, in particular for $\bm h$. The two measures of the accuracy indices are shown in Table~\ref{tb:accuracy} for each network and estimation method. The two indices took the same value in the case of the likelihood maximization \cite{Tang2008,Yeh2010}. In the case of the pseudo-likelihood maximization, the two accuracy indices were slightly different from each other, and both took approximately the same values as those derived from the maximum likelihood. In the case of the minimum probability flow, $r$ was substantially smaller than the values for the likelihood or pseudo-likelihood maximization. In contrast, $I_2/I_N$ exceeded unity because $S_1 > S_N > S_2$ for the minimum probability flow method.
\subsection{Disconnectivity graphs}
\begin{figure}[p]
\centering\includegraphics[width=5.0in]{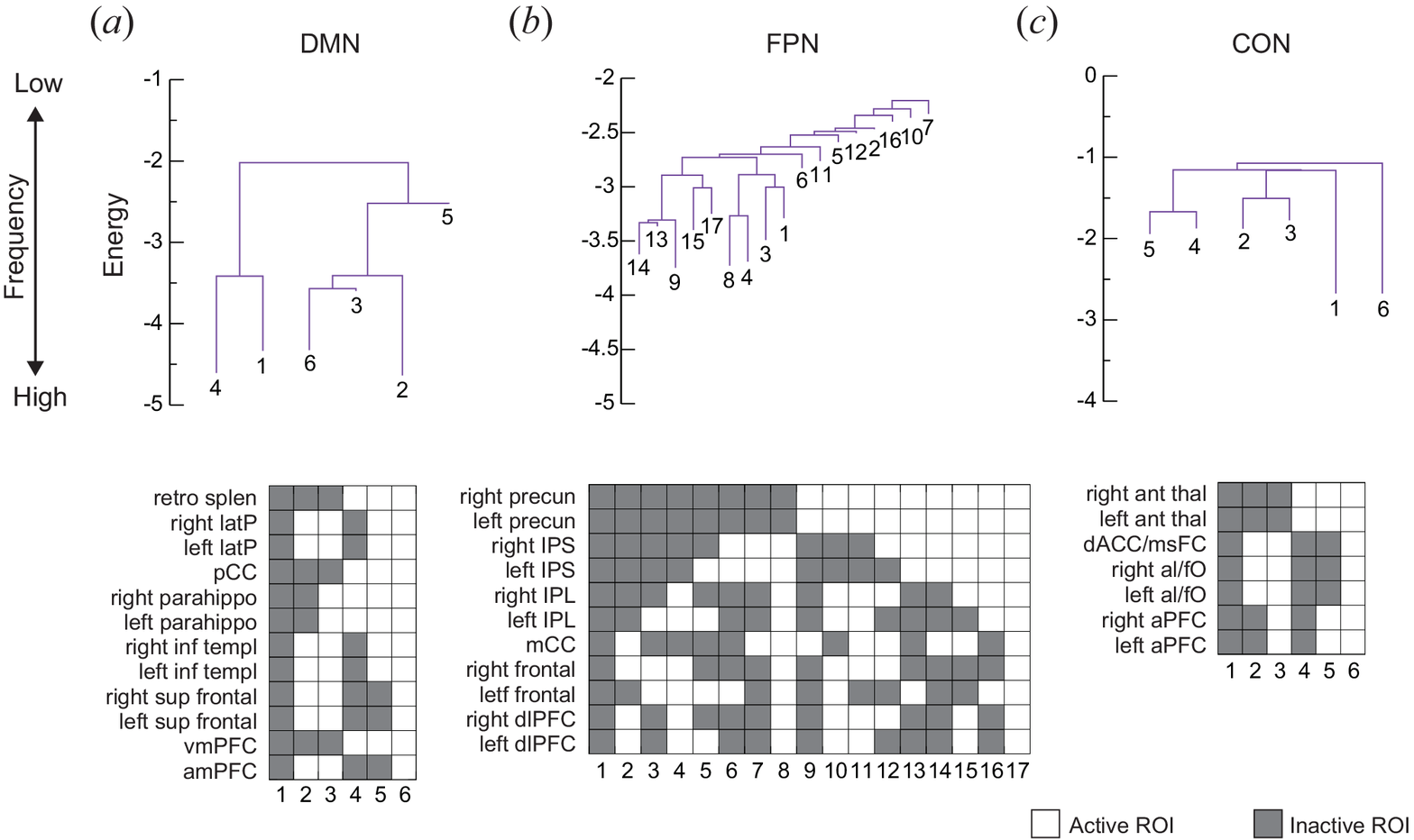}
\caption{Disconnectivity graphs for (a) DMN, (b) FPN, and (c) CON. The activity pattern at each local minimum is also shown. 
Retro splen: retro splenial cortex, latP: lateral parietal cortex, pCC: posterior cingulate cortex, parahippo: parahippocampal cortex, inf templ: inferior temporal cortex, sup frontal: superior frontal cortex, vmPFC: ventromedial prefrontal cortex, amPFC: anteromedial  prefrontal cortex, precun: precuneus, IPS: intraparietal sulcus, IPL: inferior parietal lobule, mCC: mid-cingulate cortex, frontal: lateral frontal cortex, dlPFC: dorsolateral prefrontal cortex, ant thal: anterior thalamus, dACC/msFC: dorsal anterior cingulate cortex/medial superior frontal cortex, al/fO: anterior insula/frontal operculum, aPFC: anterior prefrontal cortex.}
\label{fig:disconnectivity}
\end{figure}

Figure \ref{fig:disconnectivity} shows the disconnectivity graph of the DMN, FPN, and CON, calculated for the parameter values estimated by likelihood maximization. The two synchronized activity patterns, i.e., the activity patterns with all ROIs being active or inactive, were local minimums.
The FPN had much more local minimums than the DMN and CON did. Although the present results are opposite to our previous results using a different data set \cite{Watanabe2014b}, the reason for the discrepancy is unclear.

\subsection{Effects of the data length}
Our experiences suggest that, as the number of ROIs, $N$, increases, the pairwise MEM seems to demand a large amount of data to realize a high accuracy.
If we use $N$ ROIs, there are $2^N$ possible activity patterns. Therefore, as we increase $N$, it is progressively more likely that many of the activity patterns are unvisited. However, the MEM assigns a positive probability to such an unvisited pattern. Even if an activity pattern $\bm\sigma$ is realized by the empirical data a few times, the empirical distribution, $P_{\rm empirical}(\bm\sigma)$, would not be reliable because it is evaluated only based on a few visits to $\bm\sigma$ (divided by $t_{\max}$). If $t_{\max}$ is much larger and $\bm\sigma$ is visited proportionally many times, then we would be able to estimate $P_{\rm empirical}(\bm\sigma)$ more accurately. This exercise led us to hypothesize that the accuracy scales as a function of $t_{\max}/2^N$.

\begin{figure}[p]
\centering\includegraphics[width=5.0 in]{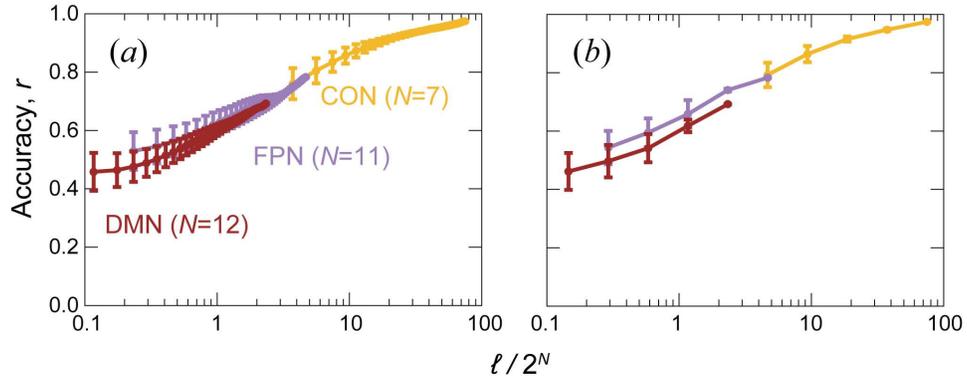}
\caption{Accuracy of the fit of the pairwise MEM as a function of the data length, $\ell$, normalized by the number of possible activity patterns, $2^N$. We obtained samples of length $\ell$ by (a) overlapping sliding time windows and (b) non-overlapping time windows. Error bars represent the standard deviation. }
\label{fig:accuracy}
\end{figure}

To examine this point, we carried out likelihood maximization on the fMRI data of varying length $\ell$ ($t_{\rm{max}}/20 \leq \ell \leq t_{\rm{max}}$) and
calculated $r$ (which coincides with $I_2/I_N$ for the maximum likelihood estimator).
For a given $\ell$, we calculated $r$ for each of the $t_{\rm{max}}-\ell$ datasets of length $\ell$, i.e., 
$\{\bm\sigma(1), \ldots, \bm\sigma(\ell)\}$, $\{\bm\sigma(2), \ldots, \bm\sigma(\ell+1)\}$, $\ldots$, $\{\bm\sigma(t_{\max}-\ell+1), \ldots, \bm\sigma(t_{\max})\}$. The average and standard deviation of $r$ as a function of $\ell/2^N$ are shown in Fig.~\ref{fig:accuracy}a for the DMN, FPN, and CON. As expected, the accuracy improved as $\ell$ increased. The results for the DMN, FPN, and CON roughly collapsed on a single curve. The figure suggests that, to achieve an accuracy of 0.8 and 0.9, each activity pattern should be visited $\approx 5$ and $\approx 16$ times on average, respectively.

Because the aforementioned sampling method used overlapping time windows to make different samples strongly depend on each other, we carried out the same test by dividing the entire time series
$\{\bm\sigma(1), \ldots, \bm\sigma(t_{\max})\}$ into two halves of length $\ell = t_{\max}/2$, four quarters of length $\ell = t_{\max}/4$, eight non-overlapping samples of length $\ell = t_{\max}/8$, and so forth. 
The results (Fig.~\ref{fig:accuracy}b) were similar to those in the case of overlapping time windows (Fig.~\ref{fig:accuracy}a).


\section{Discussion}

We explained a set of computational methods to estimate the pairwise MEM and energy landscapes from resting-state fMRI data. Novel components, as compared with our previous methods \cite{Watanabe2013,Watanabe2014a, Watanabe2014b,Watanabe2014c}, were the pseudo-likelihood maximization, the minimum probability flow, and a variant of the Dijkstra method to calculate the disconnectivity graph. We applied the methods to fMRI data collected from healthy participants and assessed the amount of data needed to secure a sufficient accuracy of fit. 

The present results suggest that the current method is admittedly demanding in terms of the amount of data, although the results should be corroborated with different data sets. In the application of the pairwise MEM to neuronal spike trains, the data length does not seem to pose a severe limit if the network size, $N$, is comparable to those in this study. This is because one typically uses a high time resolution to ensure that there are no multiple spikes within a time window (e.g., 2 ms \cite{Yu2008}, 10 ms \cite{Shlens2006}, 20 ms \cite{Schneidman2006,Tang2008,Ganmor2011a}). Then, the number of data points, $t_{\max}$, is typically much larger than in typical fMRI experiments. In fMRI experiments, the interval between two measurements, called the repetition time (TR), is typically 2--4 s, and a participant in the resting state (or a particular task condition) can be typically scanned for 5--15 mins. Then, we would have $t_{\max} = $75--450, with which we can reliably estimate the pairwise MEM model up to $N\approx 5$ (Fig.~\ref{fig:accuracy}), which is small. If we pool fMRI data from 10 participants belonging to the same group to estimate one MEM, we would have $t_{\max}=$750--4500, accommodating $N\approx 8$. This is an important limitation of our approach. Currently we cannot apply the method to relatively large brain systems (i.e., those with a10 larger number of ROIs), let alone voxel-based data.

We demonstrated the methods with fMRI data obtained from healthy participants. The same methods can be applied to different conditions of human participants including the case of medical applications, the topic of the present theme issue. Various neuropsychiatric disorders have been suggested to have dynamical footprints in the brain \cite{Rolls2011,Uhlhaas2012,Kopell2014,Wang2014a}.
Altered dynamics in the brain at various spatial and time scales may result in deformation of energy landscapes as compared with healthy controls. 


\vskip6pt

\section*{Materials and Methods}

\subsection*{Data and participants}

We used resting-state fMRI data publicly shared in the Human Connectome Project (acquisition Q10 in release S900 of the WU-Minn HCP data) \cite{VanEssen2012}. The data were collected using a 3T MRI (Skyra, Siemens) with an echo planar imaging (EPI) sequence (TR, 0.72 s; TE, 33.1 ms; 72 slices; 2.0 mm isotopic; field of view, $208 \times 180$ mm) and T1-weighted sequence (TR, 2.4 s; TE, 2.14 ms; 0.7 mm isotopic; field of view, $224 \times 224$ mm). The EPI data were recorded in four runs ($\approx 15$ min/run) while participants were instructed to relax while looking at a fixed cross mark on a dark screen.

We used such EPI and T1 images recorded from two adult participants (one female; 22-25 years old), because the amount of the data was sufficiently large for the current analysis.

\subsection*{Preprocessing and extraction of ROI data}

We preprocessed the EPI data in essentially the same manner as the conventional methods that we previously used for resting-state fMRI data \cite{Watanabe2013,Watanabe2015,Watanabe2015a} with SPM12 (www.fil.ucl.ac.uk/spm). Briefly, after discarding the first five images in each run, we conducted realignment, unwarping, slice timing correction, normalization to the standard template (ICBM 152), and spatial smoothing (full-width at half maximum $= 8$ mm). Afterwards, we removed the effects of head motion, white matter signals, and cerebrospinal fluid signals by a general linear model. Finally, we performed temporal band-pass filtering (0.01-0.1 Hz) and obtained resting-state whole-brain data. 

We then extracted a time series of fMRI signals from each ROI. The ROIs were defined as 4 mm spheres around their center whose coordinates were determined in a previous study \cite{Fair2009}. The signals at each ROI were those averaged over the sphere. In total, we obtained time-series data of length $t_{\max} = 9560$ at 30 ROIs (12 in the DMN, 11 in the FPN, and 7 in the CON).

\section*{Acknowledgment}TW acknowledges the support provided through European Commission. MO acknowledges the support provided through JSPS KAKENHI Grant No.~15H03699. NM acknowledges the support provided through JST, CREST and JST, Erato, Kawarabayashi Large Graph Project, and EPSRC Institutional Sponsorship to the University of Bristol.
Data were provided in part by the Human Connectome Project, WU-Minn Consortium (Principal Investigators: David Van Essen and Kamil Ugurbil; 1U54MH091657) funded by the 16 NIH Institutes and Centers that support the NIH Blueprint for Neuroscience Research; and by the McDonnell Center for Systems Neuroscience at Washington University.









|
\setstretch{1.0}

\newpage

\newpage

\newpage

\newpage

\clearpage
\newpage

\begin{table}[p]
\begin{center}
\begin{tabular}{ccccccc} \hline
 & \multicolumn{2}{c}{DMN} & \multicolumn{2}{c}{FPN} & \multicolumn{2}{c}{CON} \\\hline
 & $r$ & $I_2/I_N$ & $r$ & $I_2/I_N$ & $r$ & $I_2/I_N$ \\\cline{2-7}
Likelihood maximization & 0.6921  & 0.6921  & 0.7830  & 0.7830 & 0.9744  & 0.9744    \\
Pseudo-likelihood maximization & 0.6921  & 0.6972  & 0.7830  & 0.7853& 0.9745  & 0.9744   \\
Minimum probability flow & 0.6480  & 0.7437 & 0.6124  & 1.2295& ---  & ---   \\
\hline
\end{tabular}
\end{center}
\caption{Accuracy of fitting for each network and estimation algorithm.}
\label{tb:accuracy}
\end{table}

\end{document}